# HIGH-PRESSURE SYNTHESIS OF ROCK SALT LiMeO$_2$–ZnO (Me = Fe$^{3+}$, Ti$^{3+}$) SOLID SOLUTIONS

P.S. Sokolov[1,2], A.N. Baranov[2], V.A. Tafeenko[2], V.L. Solozhenko[1*]

[1] *LSPM–CNRS, Université Paris Nord, 93430 Villetaneuse, France*

[2] *Chemistry Department, Moscow State University, 119991 Moscow, Russia*

**Abstract**

*Metastable LiMeO$_2$–ZnO (Me = Fe$^{3+}$, Ti$^{3+}$) solid solutions with rock salt crystal structure have been synthesized by solid state reaction of ZnO with LiMeO$_2$ complex oxides at 7.7 GPa and 1350-1450 K. Structure, phase composition, thermal stability and thermal expansion of the recovered samples have been studied by X-ray diffraction with synchrotron radiation. At ambient pressure rock salt LiMeO$_2$–ZnO solid solutions are kinetically stable up to 670-800 K depending on the composition.*

*\*e-mail: vladimir.solozhenko@univ-paris13.fr*

**Introduction**

Zinc oxide belongs to the family of wide-band-gap semiconductors and is characterized by high exciton binding energy (60 meV) [1]. At ambient conditions ZnO has hexagonal wurtzite ($w$) structure (space group P6$_3$mc) which transforms into rock salt ($rs$) one (space group Fm3m) at pressures above 9 GPa [2]. High-pressure rock salt phase of ZnO cannot be quenched down to ambient conditions, however, metastable $rs$-Me$^{II}$O-ZnO solid solutions (Me$^{II}$ – Ni$^{2+}$, Fe$^{2+}$, Co$^{2+}$, Mn$^{2+}$) of different stoichiometry have been recently synthesized by quenching from 4.6-7.7 GPa and 1450-1650 K [3-5]. For instance, single-phase $rs$-Ni$_{1-x}$Zn$_x$O solid solutions show remarkable stability at ambient conditions in a wide concentration range (0< x ≤0.8) [3]. Instead of Me$^{II}$O monoxides, the LiMeO$_2$ complex oxide where charge compensation is attained by the combination of Li$^+$ (ionic radius is 0.76 Å) and a three-valent Me$^{3+}$ cations such as Fe$^{3+}$ (0.64 Å) and Ti$^{3+}$ (0.67 Å) [6] probably can be used. α-LiFeO$_2$ and LiTiO$_2$ are stoichiometric phases and have rock salt structure with lattice parameters 4.158 Å [7] and 4.14 Å [8], respectively, that are close to the lattice parameter of $rs$-ZnO (4.28 Å [2]).

Here we report the first synthesis of the metastable $rs$-LiMeO$_2$–ZnO solid solutions (Me = Fe$^{3+}$, Ti$^{3+}$) with high ZnO content.

**Experimental**

Commercially available powders of ZnO (99.99%, Alfa Aesar) and α-LiFeO$_2$ (99%, Aldrich) have been used as received. LiTiO$_2$ has been synthesized by hydrothermal reaction of TiO$_2$ powder with saturated aqueous LiOH solution as described in [8]. X-ray diffraction studies (G3000 TEXT Inel, Cu$K_{α1}$ radiation) have shown that all starting compounds were single phases. LiMeO$_2$–ZnO mixtures with ZnO molar fraction ($x$) from 0 to 0.9 (with 0.1 step) have been thoroughly ground in a mortar, pressed into pellets and placed into gold capsules. Quenching experiments have been performed in a toroid-type high-pressure apparatus [9] at LSPM–CNRS. Details of the experiments and pressure–temperature calibration have been described earlier [3]. Samples were gradually compressed up to 7.7 GPa, heated at desired temperature for 10-15 min, then quenched by switching off the power and slowly decompressed.

Structure of the as-synthesized solid solutions has been studied by X-ray diffraction with synchrotron radiation ($\lambda = 0.14757$ Å) at beamline BW5, HASYLAB-DESY. FullProf software [10] has been used for profile analysis and refinement procedure; the details of data processing are described elsewhere [11]. The powder diffraction patterns have been analyzed using the Le Bail method [12] to obtain the best values of lattice parameters. *In situ* high-temperature X-ray diffraction measurements ($\lambda = 0.65147$ Å) have been performed upon stepwise heating up to 1100 K at beamline B2, HASYLAB-DESY.

Thermogravimetric analysis (TG) in the 300-1200 K range has been carried out using Perkin Elmer Pyris-Diamond thermoanalyzer at heating rate of 10 K min$^{-1}$.

Microstructure morphology of the recovered samples has been studied using a high resolution scanning electron microscope LEO Supra 50 VP (Karl Zeiss) with SE2 and In-Lens secondary electron detectors.

**Results and discussion**

At ambient pressure solubility of ZnO in *rs*-LiFeO$_2$ has been studied at 1170 K (8 hour heating) for ZnO molar fraction $x \leq 0.4$. According to X-ray diffraction data, the single-phase rock salt solid solutions are forming only for $x \leq 0.2$, while at higher $x$ the mixture of rock salt and wurtzite phases is observed. Thus, at ambient pressure the solubility of ZnO in *rs*-LiFeO$_2$ does not exceed 0.2 molar fractions. Since at pressures above 5 GPa *rs*-ZnO is thermodynamically stable and forms continuous solid solutions with Me$^{II}$O (Me$^{II}$ – Ni$^{2+}$, Fe$^{2+}$, Co$^{2+}$, Mn$^{2+}$) [4,5], one may expect that ZnO solubility in LiFeO$_2$ (and LiTiO$_2$) should increase at high pressures. In order to justify this hypothesis, we have performed a systematic study of ZnO solubility in rock salt LiMeO$_2$ complex oxides at high pressures and temperatures.

A number of *rs*-LiMeO$_2$–ZnO solid solutions have been synthesized at 7.7 GPa in the 1350-1450 K temperature range. The recovered LiFeO$_2$–ZnO samples are black, whereas LiTiO$_2$–ZnO samples are white. All synthesized solid solutions are kinetically stable at ambient conditions (at least one year) and chemically stable in air (i.e. non-hygroscopic) in contrast to previously reported *rs*-(LiCrO$_2$)$_{0.33}$ZnO$_{0.67}$ solid solution [13]. According to scanning electron microscopy observations, the recovered samples are dense non-porous sintered bodies of high homogeneity (without inclusions of secondary phases or traces of the melt) with grain size of about 10-20 μm.

Fig. 1 shows a characteristic powder X-ray diffraction pattern of the *rs*-(LiFeO$_2$)$_{0.3}$(ZnO)$_{0.7}$ solid solution synthesized at 7.7 GPa and 1450 K. Le Bail analysis has shown that the recovered sample is single phase, and all observed reflections can be indexed in NaCl-type crystal structure (Fm3m space group, $a$ = 4.2430(5) Å, $R_p$ = 0.03) with cations randomly distributed allover the cationic sublattice in spite of charge difference, as it follows from the absence of superstructure diffraction lines. Previously similar disordering in the mutual positions of cations was reported for rock salt solid solutions of the LiFeO$_2$–MgO [7] and LiCrO$_2$–ZnO [13] systems.

At 7.7 GPa the upper solubility limit of ZnO in the α-LiFeO$_2$ was found to be 0.8 molar fractions i.e. under pressure the concentration range of existence of *rs*-LiFeO$_2$–ZnO solid solution is expanded by four times. At higher ZnO content ($x$ > 0.8), all recovered samples were two-phase mixtures of rock salt and wurtzite LiFeO$_2$–ZnO solid solutions. The same result has been obtained for the LiTiO$_2$–ZnO system.

In the case of Me$^{II}$O monoxides, the widest (0.3 ≤ $x$ ≤ 0.8) composition range of the existence of individual rock salt solid solutions has been observed for the NiO–ZnO system [3]. Rock salt solid solutions with FeO, CoO and MnO can be quenched down to ambient conditions with much lower ZnO content [3]. Thus, the high concentration of ZnO in the metastable *rs*-LiMeO$_2$–ZnO solid solutions can be explained only by crucial role of Li$^+$ cation in stabilization of the rock salt structure. It is known that Li$^+$ cation has strong energetic preference for the octahedral position in comparison with the tetrahedral one [14]. Besides, according to the Mooser-Pearson classification [15], significantly lower electronegativity of lithium (0.98 [16]) as compared to that of zinc (1.65 [16]) favors an octahedral rather than a tetrahedral coordination in normal valence complex oxides. At the same time three-charged cations are required to keep the charge balance in the system, thus, playing only a supplementary role.

Lattice parameters of all synthesized $rs$-LiMeO$_2$–ZnO solid solutions perfectly follow the linear concentration dependence (Vegard's law) from $a_{rs\text{-ZnO}} = 4.280$ Å [2] down to the lattice parameter of the corresponding complex oxide (inset, Fig. 1) which points to the formation of ideal substitution solid solutions.

Thermal stability of $rs$-LiMeO$_2$–ZnO solid solutions at ambient pressure has been studied by high-temperature X-ray diffraction with synchrotron radiation. Fig. 2 shows diffraction patterns of the $rs$-(LiFeO$_2$)$_{0.4}$(ZnO)$_{0.6}$ solid solution taken at different temperatures in the course of stepwise heating. Below 770 K only reflections of the pristine solid solution are observed. At higher temperatures, intensities of diffraction lines of the rock salt phase drastically decrease, while reflections of a new wurtzite phase appear, which is indicative of the decomposition of the rock salt LiFeO$_2$–ZnO solid solution into mixture of $w$-ZnO-based and $rs$-LiFeO$_2$-based solid solutions. Since thermogravimetric studies showed no weight loss upon heating up to 1200 K, one can conclude that all LiFeO$_2$-based solid solutions are thermally stable over the studied temperature range in contrast to pure LiFeO$_2$ [17].

The onset temperature of decomposition ($T_d$) of all rock salt LiMeO$_2$–ZnO solid solutions decreases with increase in ZnO content (for instance, in case of the LiFeO$_2$–ZnO system, from 770(5) K for $x = 0.6$ down to 670(5) K for $x = 0.8$). On the other hand, the nature of Me$^{3+}$ cation also influences the thermal stability, for instance, $rs$-(LiTiO$_2$)$_{0.2}$(ZnO)$_{0.8}$ is more stable ($T_d \approx 800$ K) than $rs$-(LiFeO$_2$)$_{0.2}$(ZnO)$_{0.8}$ ($T_d \approx 670$ K).

Lattice parameters of $rs$-LiFeO$_2$–ZnO solid solutions at different temperatures have been determined by high-temperature X-ray diffraction with synchrotron radiation. The temperature dependencies of the unit cell volumes of $rs$-LiFeO$_2$–ZnO solid solutions are presented in Fig. 3. For all stoichiometries, between 298 K and $T_d$ unit cell volume changes nonlinearly with temperature and can be fitted to the $V(T)=V_0[1+\alpha_1(T-298)+\alpha_2(T-298)^2]$ equation. Coefficients $\alpha_1$ and $\alpha_2$ are summarized in the Table 1. It can be seen that $\alpha_1$ monotonically increases with ZnO molar fraction ($x$), while $\alpha_2$ decreases.

**Conclusions**

Thus, rock salt $LiMeO_2$–ZnO solid solutions (Me = $Fe^{3+}$, $Ti^{3+}$) synthesized at 7.7 GPa can be quenched down to ambient conditions in the wide (from 0 to 0.8 ZnO molar fraction) concentration range. All recovered solid solutions are kinetically stable at ambient pressure up to 670-800 K depending on composition and type of $Me^{3+}$ cation.

**Acknowledgements**

The authors are grateful to T. Chauveau, K.S. Napolskii and L.A. Trusov for help with X-ray diffraction, TG and SEM studies, respectively, to J. Bednarcik for assistance with experiments at beamlime BW5, HASYLAB-DESY and to V.A. Lebedev for hydrothermal synthesis of $LiTiO_2$. X-ray diffraction measurements at B2 beamline have been carried out during beamtime allocated to the Project I-20070033 EC at HASYLAB-DESY and have received funding from the European Community's Seventh Framework Programme (FP7/2007-2013) under grant agreement No 226716. PSS is grateful to the French Ministry of Foreign Affairs for funding provided (BGF fellowship No 2007 1572).

**References**

1. C. Klingshirn, Phys. status solidi B 244 (2007) 3027-3073.
2. C.H. Bates, W.B. White, R. Roy, Science 137 (1962) 993.
3. A.N. Baranov, P.S. Sokolov, O.O. Kurakevich, V.A. Tafeenko, D. Trots, V.L. Solozhenko, High Press. Res. 28 (2008) 515-519.
4. P.S. Sokolov, A.N. Baranov, C. Lathe, V.L. Solozhenko, High Press. Res. 30 (2010) 39-43.
5. P.S. Sokolov, A.N. Baranov, C. Lathe, V.Z. Turkevich, V.L. Solozhenko, High Press. Res. 31 (2011) DOI: 10.1080/08957959.2010.521653
6. R.D. Shannon, Acta Crystallogr. A 32 (1976) 751-767.
7. G.A. Waychunas, W.A. Dollase, C.R. Ross II, Am. Mineral. 79 (1994) 274-288.
8. D.R. Zhang, H.L. Liu, R.H. Jin, N.Z. Zhang, Y.X. Liu, Y.S. Kang, J. Ind. Eng. Chem. 13 (2007) 92-96.
9. L.G. Khvostantsev, V.N. Slesarev, V.V. Brazhkin, High Press. Res. 24 (2004) 371-383.
10. J. Rodriguez-Carvajal, Physica B 192 (1993) 55-69.
11. S.I. Gurskii, V.A. Tafeenko, A.N. Baranov, Rus. J. Inorg. Chem. 53 (2008) 111-116.
12. A. Le Bail, H. Duroy, J.L. Fourquet, Mat. Res. Bull. 23 (1988) 447-452.
13. K.W. Hanck, H.A. Laitimen, J. Inorg. Nucl. Chem. 33 (1970) 63-73.
14. V.S. Urusov, Phys. Chem. Minerals. 9 (1983) 1-5.
15. E. Mooser, and W.B. Pearson, Acta Cryst. 12 (1959) 1015-1022.
16. A.L. Allred, J. Inorg. Nucl. Chem. 17 (1961) 215-221.
17. G. Bandyopadhyay, and R.M. Fulrath, J. Am. Ceram. Soc. 57 (1974) 483-486.

**Table 1**

Parameters of the $V(T) = V_0[1+\alpha_1(T-298)+\alpha_2(T-298)^2]$ equation used for fitting the thermal expansion data of rock salt $LiMeO_2$–$ZnO$ solid solutions

| Composition | $V_0$ (Å$^3$) | $\alpha_1 \times 10^5$ (K$^{-1}$) | $\alpha_2 \times 10^8$ (K$^{-2}$) |
|---|---|---|---|
| $(LiFeO_2)_{0.2}(ZnO)_{0.8}$ | 77.232(7) | 4.3(6) | 0.5(3) |
| $(LiFeO_2)_{0.3}(ZnO)_{0.7}$ | 76.402(8) | 3.5(1) | 2.5(4) |
| $(LiFeO_2)_{0.4}(ZnO)_{0.6}$ | 75.74(1) | 3.3(1) | 3.2(4) |
| $(LiTiO_2)_{0.2}(ZnO)_{0.8}$ | 76.867(5) | 4.63(5) | 0.7(1) |

**Figure 1**

Experimental (crosses), calculated (solid line) and difference (lower) X-ray diffraction pattern ($\lambda = 0.14757$ Å) of the rock salt $(LiFeO_2)_{0.3}(ZnO)_{0.7}$ solid solution quenched from 7.7 GPa and 1450 K. Insert: Lattice parameters of *rs*-LiMeO$_2$-ZnO solid solutions versus ZnO molar fraction at normal conditions; Me = $Ti^{3+}$ (squares); $Fe^{3+}$ (triangles). Error bars are smaller than the data symbols. Open symbols are the literature data [2,7,8]

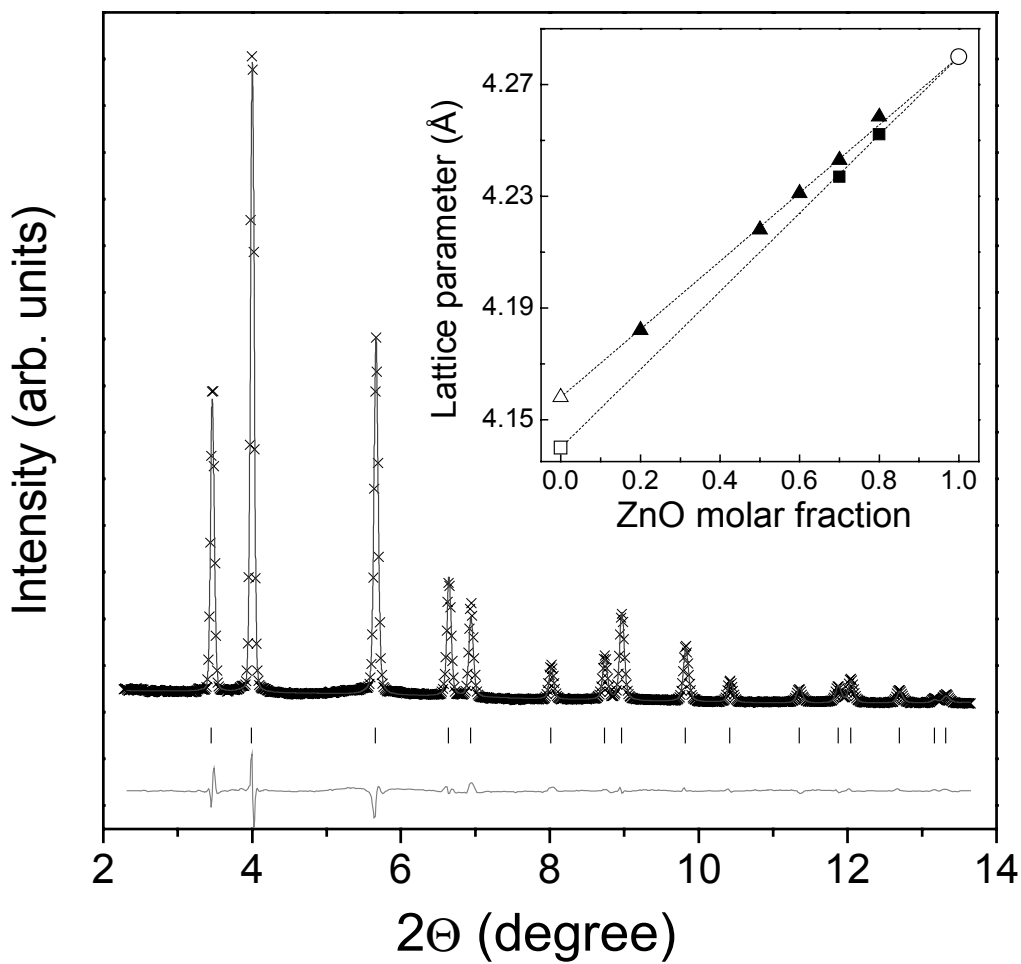

**Figure 2**

X-ray diffraction patterns ($\lambda = 0.65147$ Å) of the *rs*-(LiFeO$_2$)$_{0.4}$(ZnO)$_{0.6}$ solid solution taken at different temperatures in the course of stepwise heating at ambient pressure. The shift of peak positions of the rock salt phase is due to thermal expansion upon heating.

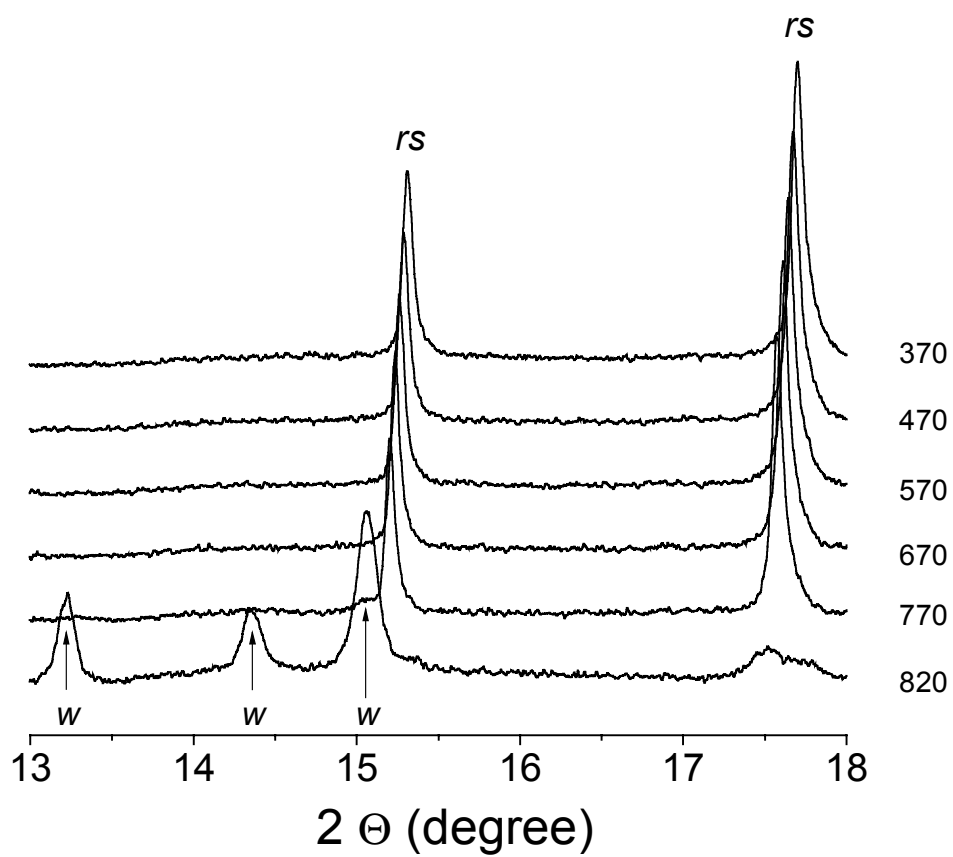

**Figure 3**

Unit cell volumes of rock salt LiFeO$_2$–ZnO solid solutions versus temperature at ambient pressure.

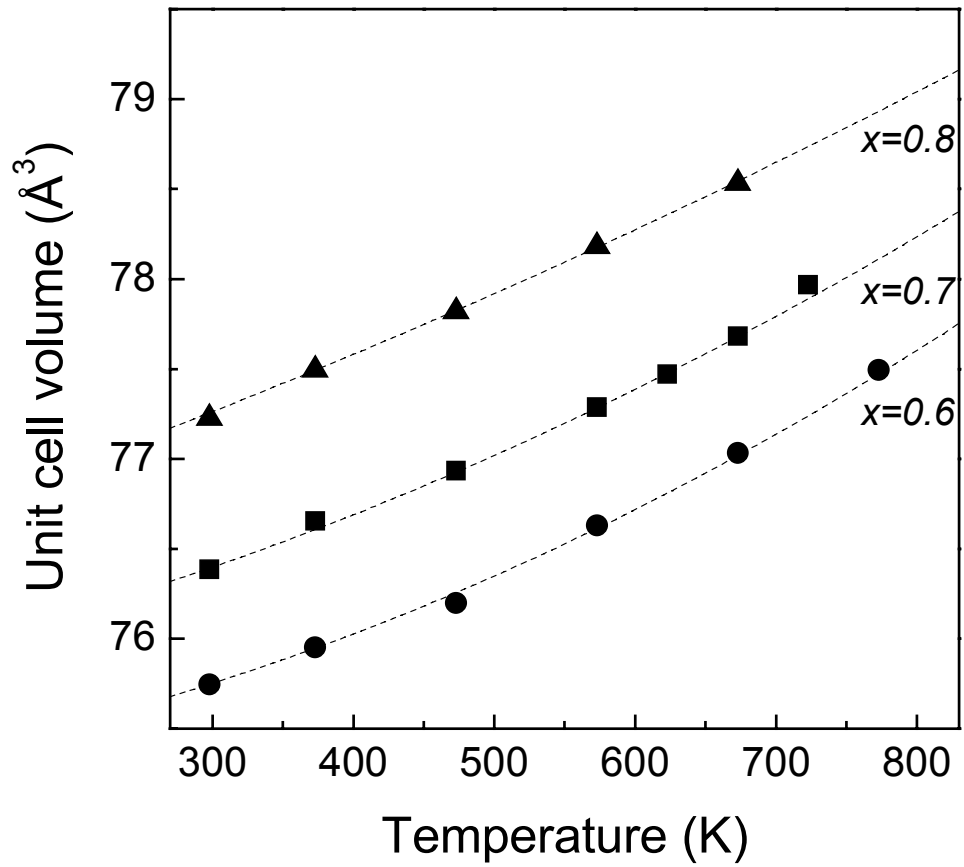